\documentclass[10pt]{article}
\usepackage{a4wide}                      % to get a wider page margen
\usepackage{epsf}                        % to make figures
\usepackage{latexsym}                    % to make fancy symbols
\usepackage{amsfonts}                    % to make fancy symbols
\usepackage{amssymb}                     % to make fancy symbols
\usepackage{mathtools}
\usepackage{mathrsfs}
\usepackage{cite}
\usepackage{float}

\numberwithin{table}{section}
\numberwithin{equation}{section}
\numberwithin{figure}{section}

\usepackage[unicode=true,colorlinks=true, urlcolor=blue, linkcolor=blue,
      citecolor=blue]{hyperref}  % to use coloured hyperlinks in text 

\def\be{\begin{equation}}
\def\ee{\end{equation}}
\def\bea{\begin{eqnarray}}
\def\eea{\end{eqnarray}}

\def\nnw{\nonumber \\ [.2cm]}

\def\hsp#1{\hspace*{#1}}

\def\part{\partial}
\def\tfrac#1#2{{\textstyle{\frac{#1}{#2}}}}
\def\half{\tfrac{1}{2}}

\def\cR{{\cal R}}

\def\cL{{\cal L}}  
\def\sqrtg{\sqrt{|g|}}

\def\mn{{\mu\nu}}
\def\mnr{{\mu\nu\rho}}
\def\mnrl{{\mu\nu\rho\lambda}}

\newcommand{\eN}{\mathrm{e}}
\newcommand{\dex}{\mathrm{d}}
\newcommand{\Dex}{\mathrm{D}}
\newcommand{\dform}[1]{\boldsymbol{#1}}
\newcommand{\dfal}{\dform{\alpha}}
\newcommand{\dfbe}{\dform{\beta}}
\newcommand{\dfk}{\dform{k}}

\newcommand{\dfL}{\dform{L}}
\newcommand{\dfQ}{\dform{Q}}
\newcommand{\dfQTrLess}{\check{\dform{Q}}}
\newcommand{\dfR}{\dform{R}}
\newcommand{\dfT}{\dform{T}}
\newcommand{\cofr}{\dform{\vartheta}}
\newcommand{\dfom}{\dform{\omega}}
\newcommand{\LCtensor}{\mathcal{E}}
\newcommand{\LCsymbol}{\varepsilon}
\newcommand{\dimM}{{n}}
\newcommand\dint[1]{#1\lrcorner}

\def\makeatletter{\catcode`\@=11}% 11:letter
\makeatletter
\def\mathbox#1{\hbox{$\m@th#1$}}%
\def\math@ccstyles#1#2#3#4#5#6#7{{\leavevmode
      \setbox0\mathbox{#6#7}%
      \setbox2\mathbox{#4#5}%
      \dimen@ #3%
      \baselineskip\z@\lineskiplimit#1\lineskip\z@
      \vbox{\ialign{##\crcr
             \hfil \kern #2\box2 \hfil\crcr
             \noalign{\kern\dimen@}%
             \hfil\box0\hfil\crcr}}}}

\def\mathaccstyles{\math@ccstyles\maxdimen}
\def\maththroughstyles{\math@ccstyles{-\maxdimen}}

\def\unity%
 {\maththroughstyles{.45\ht0}\z@\displaystyle {\mathchar"006C}\displaystyle 1}
\begin{document}

\rightline{\today}
{~}
\vspace{1.4truecm}

%%%%%%%%%%%%%%%%%
\centerline{\LARGE \bf On the topological character of  metric-affine}
\vspace{.3truecm}
\centerline{\LARGE \bf Lovelock Lagrangians in critical dimensions}
\vspace{1.5truecm}

\centerline{
    {\bf Bert Janssen and}
    {\bf Alejandro Jim\'enez-Cano}}

\vspace{.4cm}
\centerline{{\it Departamento de F\'isica Te\'orica y del Cosmos and}}
\centerline{{\it Centro Andaluz de F\'isica de Part\'iculas Elementales}}
\centerline{{\it Facultad de Ciencias, Avda Fuentenueva s/n,}}
\centerline{{\it Universidad de Granada, 18071 Granada, Spain}}
\centerline{E-mails: {\tt bjanssen@ugr.es}, \ {\tt alejandrojc@ugr.es}}
\vspace{2cm}

%%%%%%%%%%%%%%%%%
\centerline{\bf ABSTRACT}
\vspace{.5cm}

\noindent
In this paper we prove that the $k$-th order metric-affine Lovelock Lagrangian  is not a total derivative in the critical dimension $\dimM=2k$ in the presence of non-trivial non-metricity. We use a bottom-up approach, starting with the study of the simplest cases, Einstein-Palatini in two dimensions and Gauss-Bonnet-Palatini in four dimensions, and focus then on the critical Lovelock Lagrangian of arbitrary order. The two-dimensional Einstein-Palatini case is solved completely and the most general solution is provided. For the Gauss-Bonnet case, we first give a particular configuration that violates at least one of the equations of motion and then show explicitly that the theory is not a pure boundary term. Finally, we make a similar analysis for the $k$-th order critical Lovelock Lagrangian, proving that the equation of the coframe is identically satisfied, while the one of the connection only holds for some configurations. In addition to this, we provide some families of non-trivial solutions. 

\newpage

%%%%%%%%%%%%%%%%%%%%%%%%%%%%%%%%%%%%%%%%%%%%%%%%%%%%%%%%%
\section{Introduction}

Lovelock gravities are a family of higher-curvature Lagrangian terms that form a natural extension to standard General Relativity. Introduced in the early 1970s by Lovelock \cite{Lovelock1, Lovelock2} (though the easiest non-trivial case, Gauss-Bonnet gravity, was already identified by Lanczos in 1938 \cite{Lanczos}), they are characterised as the unique higher-curvature terms that give rise to second-order differential equations after varying with respect to the spacetime metric. It is precisely this property that makes the theory such a natural extension, as it is guaranteed to be ghost-free \cite{Zwiebach, Zumino}. In this way, Lovelock gravities are singled out with respect to all other higher-curvature extensions, which generically do suffer this problem. In addition, Lovelock gravities appear as string corrections to supergravity \cite{CHSW, GW, GZ, Tseytlin, PZ, MT} and over the years have attracted a lot of attention in cosmology and string theory as corrections to  black hole solutions and FRW models, as alternatives of dark matter or dark energy and to obtain corrections to holographic models (see for example \cite{BB, CNO, NOS, BEMPSS, BKP, CEGG, DDMM}). 

In $\dimM$ dimensions, the (metric) Lovelock action is defined as
\begin{equation}
\mathring{S}_{\rm Lov}^{(\dimM)} \ = \ \sum_{k=1}^{\lfloor\dimM/2\rfloor} \mathring{S}_k^{(\dimM)}  \ = \ \int \dex^\dimM x \ \sqrtg \ \sum_{k=1}^{\lfloor\dimM/2\rfloor}\lambda_k \mathring{{\mathcal{L}}}_k^{(\dimM)} \,.
\end{equation}
Here $\lfloor x\rfloor$ is the floor function, $\lambda_k$ are certain dimensionful constants and the Lagrangian densities $\mathring{\mathcal{L}}_k^{(\dimM)}$ are given by
\bea
\mathring{\mathcal{L}}_k^{(\dimM)} \ = \ \frac{(2k)!}{2^{k}}\ 
\delta^{\mu_1\nu_1... \mu_k\nu_k}_{\alpha_1\beta_1... \alpha_k \beta_k} \, 
g^{\alpha_1 \rho_1}...\, g^{\alpha_k \rho_k} 
\mathring{\cR}_{\mu_1\nu_1 \rho_1}{}^{\beta_1} \, ... \,
\mathring{\cR}_{\mu_k\nu_k \rho_k}{}^{\beta_k}\,,
\label{Lovelock-LC}
\eea
where $\mathring{\cR}_{\mu\nu\rho}{}^\lambda$ is the Riemann tensor constructed with the Levi-Civita connection $\mathring\Gamma_\mn{}^\rho$,
\bea
\mathring\Gamma_{\mn}{}^\rho &=&
\half g^{\rho\lambda}\Bigl[\partial_\mu g_{\lambda\nu} \ + \ \partial_\nu g_{\mu\lambda}
      \ - \ \partial_\lambda g_\mn\Bigr]\,, 
\\ [.2cm]
\mathring{\cR}_{\mu\nu\rho}{}^\lambda 
&=& \partial_\mu \mathring\Gamma_{\nu\rho}{}^\lambda
\ - \ \partial_\nu \mathring\Gamma_{\mu\rho}{}^\lambda
\ + \ \mathring\Gamma_{\mu\sigma}{}^\lambda \, \mathring\Gamma_{\nu\rho}{}^\sigma
\ - \  \ \mathring\Gamma_{\nu\sigma}{}^\lambda \, \mathring\Gamma_{\mu\rho}{}^\sigma \,.
\label{LCdef}
\eea
and the multi-index delta represents the antisymmetrised product of Kronecker deltas,
\bea
\delta^{\mu_1\nu_1\dots\mu_k\nu_k}_{\alpha_1\beta_1\dots \alpha_k\beta_k}\,
&= & \delta^{[\mu_1}_{\alpha_1} \, \delta^{\nu_1}_{\beta_1} \dots
       \delta^{\mu_k}_{\alpha_k} \, \delta^{\nu_k]}_{\beta_k} \\ [.2cm]
   &=& \tfrac{1}{(2k)! (\dimM-2k)!} \ \mathrm{sgn}{(g)} \ |g| \ \LCsymbol^{\mu_1\nu_1\dots\mu_k\nu_k\sigma_1 \dots \sigma_{\dimM-2k}}\
       \LCsymbol_{\alpha_1\beta_1\dots \alpha_k\beta_k\sigma_1 \dots \sigma_{\dimM-2k}},
       \nonumber
\eea
with $\LCsymbol_{\mu_1 \dots \mu_\dimM} = \dimM! \delta^1_{[\mu_1} ... \delta^\dimM_{\mu_\dimM]}$ the totally alternating Levi-Civita symbol.\footnote{ In our signature convention $(+\,-\,...\,-)$, we have $\mathrm{sgn}{(g)}=(-1)^{\dimM-1}$.}

In this paper we focus our study on the dynamical properties of each of these terms separately,
\begin{equation} 
\mathring{S}_k^{(\dimM)} \ = \ \lambda_k \int \dex^\dimM x \ \sqrtg \ \mathring{{\mathcal{L}}}_k^{(\dimM)} \label{Lovelock-LC-action} \,.
\end{equation}
From now on, we will refer to $\mathring{S}_k^{(\dimM)}$ as the $k$-th order (metric) Lovelock term. Working out the lowest order cases, it is easy to see that the first and second order Lovelock terms are the Einstein-Hilbert action and the Gauss-Bonnet term respectively,
\bea
&& \mathring{S}_1^{(\dimM)} \ = \ \lambda_{\rm EH} \int \dex^\dimM x \,\sqrtg \, \mathring\cR \ ,\nnw
&& \mathring{S}_2^{(\dimM)} \ = \ \lambda_{\rm GB} \, \int \dex^\dimM x \, \sqrtg \, \Bigl[ \mathring\cR^2
  \, - \, 4\mathring\cR_\mn\mathring\cR^\mn
  \, + \, \mathring\cR_\mnrl\mathring\cR^\mnrl\Bigr]\, ,
\eea
where $\lambda_{\rm EH}= (2\kappa)^{-1}$ and the Ricci tensor and scalar are defined as $\mathring\cR_\mn = \mathring\cR_{\mu\lambda\nu}{}^\lambda$ and $\mathring\cR = g^\mn \mathring\cR_\mn$.

The dynamical properties of each Lovelock term depend crucially on the number $\dimM$ of dimensions in which the theory is formulated. From the definition it is clear that the $k$-th order Lovelock term vanishes identically in any dimension $\dimM<2k$. Also it is well known \cite{Zumino} that in $\dimM=2k$ it is a topological term, proportional to the $2k$-dimensional Euler characteristic and hence does not contribute to the equations of motion. Only in $\dimM> 2k$ the $k$-th order term is dynamical and yields non-trivial physics.

The proof that Lovelock terms in critical dimensions (i.e. the $k$-th order term in $\dimM=2k$) are proportional to the Euler characteristic is traditionally done via the generalised Gauss-Bonnet Theorem (see for example \cite{EGH, Nakahara} for a pedagogical introduction). Indeed, the action can be written as a surface term, i.e. as the integral of a total derivative $S_k = \int \dex^{2k}x \, \partial_\mu F^\mu$ of some functions $F^\mu (g, \partial g)$ \cite{YP}.

An interesting question is how much of this picture remains true if one abandons the traditional Riemannian geometry and allows for general affine connections $\Gamma_{\mn}{}^\rho$. Remember that in differential geometry the metric $g_\mn$ and the affine connection $\Gamma_{\mn}{}^\rho$ are in principle two independent variables, that describe different geometrical properties of the manifold $M$: the metric measures distances between points and angles between vectors in the tangent space $T_p(M)$ of a given point $p$, while the affine connection defines parallel transport of vectors between tangent spaces and hence determines the curvature of the manifold. Only in Riemannian geometry, the affine connection is chosen to be the Levi-Civita connection (\ref{LCdef}), which is a function exclusively of the metric. Therefore, the geometrical properties of the manifold are completely determined by $g_\mn$. On the other hand, in general metric-affine gravities, the affine connection $\Gamma_{\mn}{}^\rho$ is considered an independent variable, with its own equations of motion, that dictate the dynamics and hence the
allowed solutions.

It is well known that within the space of affine connections, the Levi-Civita connection is identified as the only one that has both vanishing torsion $T_{\mn}{}^\rho = \Gamma_{\mn}{}^\rho - \Gamma_{\nu\mu}{}^\rho$ and vanishing non-metricity $Q_\mnr = -\nabla_\mu g_{\nu\rho}$. In metric-affine gravity, the extra degrees of freedom come therefore from non-trivial torsion, non-metricity, or both. The aim of this paper is to study how these additional degrees of freedom affect the dynamical properties of the Lovelock terms discussed above. In particular, whether metric-affine Lovelock gravities in critical dimensions maintain their  topological character, or whether the torsion and/or non-metricity give rise to non-trivial dynamics.

The metric-affine Lovelock terms (and hence the total metric-affine Lovelock action) are defined in an analogous way as their metric counterparts (\ref{Lovelock-LC})-(\ref{Lovelock-LC-action}), but using a general connection,
\bea
&& S_k^{(\dimM)} \ = \ \lambda_k \int \dex^\dimM x \ \sqrtg \ {\mathcal{L}}_k^{(\dimM)}  \,,
 \label{Lovelock-Palatini2} \\ [.2cm]
&& \mathcal{L}_k^{(\dimM)} \ =\ \frac{(2k)!}{2^{k}}\ 
\delta^{\mu_1\nu_1... \mu_k\nu_k}_{\alpha_1\beta_1... \alpha_k \beta_k} \, \,
g^{\alpha_1 \rho_1}... g^{\alpha_k \rho_k}
\cR_{\mu_1\nu_1 \rho_1}{}^{\beta_1}(\Gamma) \, ... \,
\cR_{\mu_k\nu_k \rho_k}{}^{\beta_k}(\Gamma) \,,
\label{Lovelock-Palatini}
\eea
where now the Riemann tensor 
$\cR_{\mu\nu\rho}{}^\lambda (\Gamma) =  \partial_\mu \Gamma_{\nu\rho}{}^\lambda
-  \partial_\nu \Gamma_{\mu\rho}{}^\lambda
+  \Gamma_{\mu\sigma}{}^\lambda\Gamma_{\nu\rho}{}^\sigma
- \Gamma_{\nu\sigma}{}^\lambda\Gamma_{\mu\rho}{}^\sigma$
is constructed from the general affine connection $\Gamma_{\mn}{}^\rho$ and has in general less symmetries than its Levi-Civita counterpart.

In principle the coefficients $\lambda_k$ are arbitrary, but considering metric-compatible connections (i.e. $Q_\mnr =0$), it has been shown in \cite{TZ} (see also \cite{CPRS}) that demanding the theory to have the maximum number of degrees of freedom, the relative coefficients are fixed in terms of the number $\dimM$ of dimensions and a parameter that can be interpreted as an (anti-)de Sitter radius. Furthermore, the coefficients are such that in odd dimensions the complete Lovelock action $S_{\rm Lov}^{(\dimM)}|_{Q=0}=\sum_k S_k^{(\dimM)}|_{Q=0}$ can be written as the Chern-Simons form of the $\dimM$-dimensional (anti-)de Sitter group and in even dimensions in a Born-Infeld-like form. We are not aware whether the same properties hold when non-trivial non-metricity $Q_\mnr \neq 0$ is included.

We insist that in the action (\ref{Lovelock-Palatini2}) both the metric and the connection are considered to be dynamical fields, each with its own equation of motion. 
Some results are known about the space of allowed connections in metric-affine Lovelock theories. In \cite{ESJ, BJB, DP} it was shown that general metric-affine Lagrangians $\cL(g_\mn, \cR_\mnr{}^\lambda)$ allow the Levi-Civita connection (\ref{LCdef}) as a solution only if the Lagrangian is Lovelock. In this sense, the metric formalism is always a consistent truncation of metric-affine Lovelock theories \cite{DP}. On the other hand, there are indications that Levi-Civita is in general not the only allowed connection: in \cite{Pons, BJJOSS} it was proven that the most general connection for the metric-affine Einstein action in $\dimM>2$ is of the form
\be
\bar \Gamma_{\mn}{}^\rho \ = \ \mathring\Gamma_{\mn}{}^\rho
\ + \ A_\mu \, \delta^\rho_\nu \,,
\label{Palatini}
\ee
for arbitrary vector fields $A_\mu$. However it was also shown that this vector field is in fact unphysical and should be interpreted as the parameter of a projective symmetry
\be
\Gamma_{\mn}{}^\rho \ \rightarrow \ \Gamma_{\mn}{}^\rho \ + \ A_\mu \delta^\rho_\nu \,,
\label{projsymm}
\ee
of the Einstein-Palatini action \cite{Eisenh, JS}. In this sense, the metric and the metric-affine formalisms are physically completely equivalent for the Einstein action in $\dimM>2$.

In \cite{JJOS, JJO} it was shown that the projective symmetry (\ref{projsymm}) is present in any metric-affine Lovelock theory (\ref{Lovelock-Palatini}) and hence that the affine connection (\ref{Palatini}) is a solution (physically equivalent to Levi-Civita) in all these cases. However it is not known whether (\ref{Palatini}) is the only allowed solution and hence whether the metric and the metric-affine formalism are also equivalent for general Lovelock theories. In fact there are indications that this is not the case. 

Recently, a first solution, physically inequivalent to (\ref{Palatini}), was presented for the $k=2$ case. To be specific, in \cite{JJO} it was shown that the projective Weyl connection\footnote{We call the projective Weyl connection the generalization of the Weyl connection $\tilde \Gamma_{\mn}{}^\rho = \mathring\Gamma_{\mn}{}^\rho   +  B_\mu \delta^\rho_\nu + B_\nu \delta_\mu^\rho- B^\rho g_\mn$ in presence of the projective symmetry (\ref{projsymm}).}
\be
\tilde \Gamma_{\mn}{}^\rho \ = \ \mathring\Gamma_{\mn}{}^\rho
\ + \ A_\mu \, \delta^\rho_\nu \ + \ B_\nu \, \delta_\mu^\rho \ - \ B^\rho \, g_\mn
\label{GBPsol}
\ee
is a solution for the pure Gauss-Bonnet-Palatini gravity for arbitrary $B_\mu$, but only in $\dimM=4$ (which is precisely the critical dimension corresponding to $k=2$). It was argued that the existence of the solution is related to the conformal invariance (i.e. invariance under rescalings of the metric) of both the metric and the metric-affine Gauss-Bonnet term in $\dimM=4$. Furthermore, it was shown that the transformation
\be
\Gamma_{\mn}{}^\rho \ \rightarrow \ \Gamma_{\mn}{}^\rho
\ + \ B_\nu \, \delta_\mu^\rho \ - \ B^\rho \, g_\mn  
\label{B-transf}
\ee
is a symmetry of the four-dimensional Gauss-Bonnet-Palatini theory when it is restricted to metric-compatible connections, $\cL_2^{(4)}|_{Q=0}$, but not of the full theory $\cL_2^{(4)}$ with arbitrary connections. Since in the full theory there is no symmetry transformation that relates (\ref{GBPsol}) to the known solutions (\ref{Palatini}), the new solution is interpreted as physically inequivalent to the Levi-Civita connection.\footnote{It is well known that symmetries of a consistently truncated theory that do not leave the full Lagrangian invariant, act as solution generating transformations in the full theory.}

The existence of a non-trivial solution and of a transformation that is a symmetry of the truncated, but not of the full Lagrangian, are the first hints that the four-dimensional Gauss-Bonnet theory might not be a total derivative in the metric-affine formulation. As these topological theories do not have dynamical equations of motion, there are no restrictions on its field content and any field configuration appears as an allowed solution. However, as this might be true for the truncated critical Gauss-Bonnet-Palatini term $\cL_2^{(4)}|_{Q=0}$, it is clearly not the case for the full theory $\cL_2^{(4)}$. The idea is then that the non-metricity $Q_\mnr$ spoils the topological character of the critical Gauss-Bonnet-Palatini term. Moreover, as the solution (\ref{GBPsol}) is conjectured to exist in all $k$-th order Lovelock terms in $\dimM=2k$ \cite{JJO}, these properties would hold for all critical Lovelock theories.\footnote{From \cite{TZ} it is clear that the torsionful metric-compatible connections still yield topological theories.} 

The aim of this paper is to proof that this is indeed the case.  In Section \ref{EH} we will deal with the simplest of all critical Lovelock theories, namely the two-dimensional Einstein term in the metric-affine formulation. We will compute the equations of motion of the metric and the affine connection and show that they do not impose any conditions on the metric or the torsion, but restrict the non-metricity in a non-trivial way. We will also show that the $\dimM=2$ Einstein-Palatini action can be written as a total derivative plus terms that depend on $Q_\mnr$.  In Section \ref{dim=4GB} we will perform a similar analysis for the four-dimensional Gauss-Bonnet-Palatini term. We will first show that the Lagrangian is not a total derivative in the presence of non-metricity. Afterwards, as this theory is too complicated to be solved in general, we present a field configuration the does not obey the equations of motion of the theory, proving that the latter impose non-trivial conditions. In Section \ref{dim=2kLovelock} we study the general critical $k$-th order Lovelock term, deriving its equations of motion, and discussing examples of field configurations that do and do not satisfy these.  We will start our discussion in the language of affine connections, but gradually move to language of differential forms, which turns out to be especially useful to treat with Lovelock Lagrangians and their equations of motion. A brief review of differential forms, general properties of connections and the derivations of the equations of motion can be found in the Appendices.

%%%%%%%%%%%%%%%%%%%%%%%%%%%%%%%%%%%%%%%%%%%%%%%%%%%%%%%%%
\section{The Einstein-Palatini action in \texorpdfstring{$\dimM=2$}{n=2}}
\label{EH}

\subsection{Solving the Einstein-Palatini theory}

The two-dimensional metric-affine Einstein-Palatini term is given by
\be
  S_1^{(2)} \ = \ \frac{1}{2\kappa} \int \dex^2x \, \sqrtg \, \delta^\mn_{\alpha\beta} \, g^{\alpha\rho} \, \cR_\mnr{}^{\beta} (\Gamma) \ = \ \frac{1}{2\kappa} \int \dex^2x \, \sqrtg \, \cR (g,\Gamma)\,,
\label{dim=2EH}                  
\ee
where $\cR(g,\Gamma) = g^\mn \cR_\mn(\Gamma) = g^\mn \cR_{\mu\lambda\nu}{}^\lambda(\Gamma)$. The equations of motion for the metric and the connection are given by \cite{BJJOSS, JJO}
\begin{align}
0 &=  \cR_{(\mn)} \ - \ \frac{1}{2} \, g_\mn \, \cR\,, \label{EH-EOM}\\
0 &=  Q_\lambda{}^\mn \ - \ Q_\sigma{}^{\sigma\nu} \, \delta^\mu_\lambda
\ - \ \frac{1}{2} \, Q_{\lambda\sigma}{}^\sigma \, g^\mn
\ + \ \frac{1}{2}\, Q^{\nu\sigma}{}_\sigma \, \delta^\mu_\lambda
 \\
  &\quad \qquad  \qquad 
\ - \ T_{\sigma\lambda}{}^\sigma \, g^\mn
\ + \ T_{\sigma\rho}{}^\sigma \, g^{\rho\nu} \, \delta^\mu_\lambda
\ + \ T_{\sigma\lambda}{}^\mu \, g^{\sigma\nu} \,. \nonumber
\end{align}
Note that the Einstein equation is automatically traceless in two dimensions and can not be further simplified. Similarly, in $\dimM=2$ only the $\delta^\lambda_\mu$
trace of the connection equation is non-trivial and relates the different traces of the non-metricity as
\be
Q_\sigma{}^{\sigma\nu} \ = \ \frac{1}{2} \, Q^{\nu\sigma}{}_\sigma \,.
\ee
Substituting this condition into (\ref{EH-EOM}), we find
\bea
Q_{\lambda\mn} \ - \ \frac{1}{2} Q_{\lambda\sigma}{}^\sigma \, g_\mn
\ + \ T_{\lambda\sigma}{}^\sigma \, g_\mn
\ - \ T_{\nu\sigma}{}^\sigma \, g_{\mu\lambda}
\ - \ T_{\lambda\nu\mu} \ = \ 0 \,.
\label{tracelessGammaEH}
\eea

The connection equation (\ref{tracelessGammaEH}) can be best solved dividing the torsion and the non-metricity into their irreducible components. As can be seen in Appendix \ref{IrrepsTQ}, in $\dimM=2$ the torsion is pure trace,
\be
T_\mn{}^\rho \ = \ 2 \, T_{[ \mu |\sigma|}{}^\sigma \, \delta_{\nu]}^\rho \,,
\label{EH-torsion}
\ee
as the other irreducible parts are identically zero. Plugging (\ref{EH-torsion}) into the connection equation (\ref{tracelessGammaEH}), it is easy to see that the torsion drops out of the equation, but that the non-metricity should obey the condition
\be
Q_{\lambda\mu\nu} \ = \ \frac{1}{2} \, Q_{\lambda\sigma}{}^\sigma \, g_\mn.
\label{EH-Q}
\ee
In other words, also the non-metricity is pure trace. The most general affine connection that satisfies both (\ref{EH-torsion}) and (\ref{EH-Q}) is the projective Weyl connection (\ref{GBPsol}),
\be
\tilde \Gamma_{\mn}{}^\rho \ = \ \mathring\Gamma_{\mn}{}^\rho \ + \ A_\mu \, \delta^\rho_\nu \ + \ B_\nu \, \delta_\mu^\rho \ - \ B^\rho \, g_\mn, \label{GBPsol2}
\ee
as the torsion and non-metricity are given by
\be
  T_{\mn}{}^\rho \ =  \ 2\, (A_{[\mu} - B_{[\mu} ) \, \delta_{\nu]}^\rho, \qquad{}  Q_\mnr \ = \ 2 A_\mu \, g_{\nu\rho}. \label{EinstSolTorNMet}
\ee
This connection (\ref{GBPsol2}) was conjectured in \cite{JJO} to be a solution of any critical Lovelock theory and indeed we find it here as the most general solution of the $\dimM=2$ metric-affine Einstein-Palatini action.

Once the most general connection is known, let us look at the Einstein equation. The two-dimensional Ricci tensor and scalar constructed from (\ref{GBPsol2}) are given by
\be
\cR_\mn(\tilde \Gamma) \ = \ \mathring \cR_\mn \ + \ F_\mn (A) \ + \ \mathring\nabla_\lambda B^\lambda \,g_\mn,
\hsp{1cm}
\cR(g,\tilde \Gamma) \ = \ \mathring \cR \ + \ 2\mathring\nabla_\lambda B^\lambda,
\ee
where $\mathring\nabla$ is the Levi-Civita covariant derivative and $F_\mn(A) = 2 \partial_{[\mu}A_{\nu]}$. It is easy to see that the metric-affine Einstein equation (\ref{EH-EOM}) with the connection on-shell reduces to the Levi-Civita one:
\be
0 \ = \  \mathring\cR_{\mn} \ - \ \frac{1}{2} \, g_\mn \, \mathring\cR\,.
\ee
The latter does not impose any conditions on the metric, as the metric Einstein-Hilbert term is a topological invariant in $\dimM=2$. Indeed, it is well known that in two dimensions all metrics are conformally flat, $g_\mn = e^{2\phi(x)}\eta_\mn$, such that the Ricci tensor and scalar,
\be
\mathring \cR_\mn \ = \ \mathring\nabla^2 \phi \, \eta_\mn,
\hsp{2cm}
\mathring \cR \ = \ 2 \, e^{-2\phi} \, \mathring\nabla^2 \phi
\ee
yield an Einstein tensor that vanishes identically.

We thus find that the two-dimensional Einstein-Palatini term leaves both the metric and the torsion completely undetermined, but puts dynamical conditions on the non-metricity. Indeed, although the pure trace conditions (\ref{EH-torsion}) and (\ref{EH-Q}) of the torsion and the non-metricity look similar, it should be clear that their origin is completely different: (\ref{EH-torsion}) is a group-theoretical argument valid in general in two dimensions, while it is the connection equation  (\ref{EH-EOM}) that forces the non-metricity to be pure trace, $Q_\mnr = Q_{\mu\sigma}{}^\sigma g_{\nu\rho}$. As can be seen in  Table \ref{tabledof}, besides the trace, the non-metricity has another 4 degrees of freedom $\check{Q}_{\mu\nu\rho}$, which are set to zero by the dynamics of the theory. We refer to Appendix \ref{IrrepsTQ} for a quick discussion about the number of degrees of freedom of the different irreducible parts of $T_{\mu\nu}{}^{\rho}$ and $Q_{\mu\nu\rho}$. A more detailed study can be found in \cite{McCrea, HMMN}.

The fact that there are non-trivial conditions on  $Q_\mnr$ strongly suggests that the two-dimen\-sio\-nal Einstein-Palatini action is topological when endowed with a torsionful metric-com\-pati\-ble connection, but not for connections with non-vanishing $Q_\mnr$. We will now show that indeed, in general, the Einstein-Palatini action (\ref{dim=2EH}) can be written as a sum of a total derivative term and a term that depends on $Q_\mnr$.

\begin{table}[t]
  \begin{center} \renewcommand{\arraystretch}{1.5}
  \begin{tabular}{|c|c|c|c|}
    \hline 
    Tensor & d.o.f. in $\dimM$ dim. & d.o.f. in 2 dim. & Condition imposed by EoM\tabularnewline
    \hline \hline 
    $T_{\mu\nu}{}^{\rho}$ & $\frac{1}{2}\dimM^{2}(\dimM-1)$ & 2 (pure trace) & None\tabularnewline
    \hline 
    $Q_{\mu\lambda}{}^{\lambda}$ & $\dimM$ & 2 & None \tabularnewline
    \hline 
    $\check{Q}_{\mu\nu\rho}$ & $\frac{1}{2}\dimM(\dimM+2)(\dimM-1)$ & 4 & They are zero\tabularnewline
    \hline 
  \end{tabular}\renewcommand{\arraystretch}{1}
    \parbox{0.95\textwidth}{
      \caption{{\it Splitting of the degrees of freedom of the affine connection in general dimension and in $\dimM=2$. The last column shows the conditions imposed by the equations of motion of the two-dimensional Einstein-Palatini theory. Observe that the indetermination of the trace of the non-metricity holds in arbitrary $\dimM$ due to projective symmetry.}}}
  \end{center}
  \label{tabledof} 
\end{table}

%%%%%%%%%%%%
\subsection{The $\dimM = 2$ Einstein-Palatini action is not a total derivative}
\label{SectDiffForms}

In order to study the form of the Einstein-Palatini action, we will quickly introduce the Vielbein formalism in the language of differential forms (we refer to Appendix \ref{AppendixDiffForms} for a quick review of differential forms). We start considering an arbitrary smooth distribution of bases $\boldsymbol{e}_a$ over the different tangent spaces, which we will call a \emph{frame}, and the dual distribution of cobases or a  \emph{coframe} $\cofr^a$,
\begin{equation}
  \boldsymbol{e}_a \, = e^\mu{}_a \, \boldsymbol{\partial}_\mu \,,
  \qquad{}\qquad{} \cofr^a \ = \ e^a{}_\mu \, \dex x^\mu\,.
\end{equation}
In other words, we have that $\cofr^a(\boldsymbol{e}_b)= e^a{}_\mu e^\mu{}_b=\delta^a_b$ (in addition, we also have $e^\nu{}_a e^a{}_\mu=\delta^\nu_\mu$). We can now obtain the components of the metric in this new basis as
\begin{equation}
g_{ab} \ = \ e^\mu{}_a \, e^\nu{}_b \, g_{\mu\nu}\,.
\end{equation}
In principle, these anholonomic components of the metric  $g_{ab}$ are completely general. However, in order to simplify many expressions, we will  we consider throughout the paper a particular $\mathrm{GL}(\dimM,\mathbb{R})$ gauge for the coframe, in which $g_{ab}$ is independent of the point, i.e. its components are constant, $\dex g_{ab} = \partial_\mu g_{ab}\dex x^\mu =0$.

In the language of differential forms, the affine degrees of freedom of the theory are encoded in the \emph{connection 1-form}, $\dfom_a{}^b = \omega_{\mu a}{}^b \dex x^\mu$, whose components are nothing else than the components of the affine connection, transformed to the anholonomic basis:\footnote{This expression is sometimes called the ``Vielbein postulate'', but in our approach it is simply the definition of $\omega_{\mu a}{}^{b}$.}
\begin{equation}
\omega_{\mu a}{}^{b} \ = \  e^{\nu}{}_{a}\, e^{b}{}_{\lambda}\, \Gamma_{\mu\nu}{}^{\lambda} \ + \ e^{b}{}_{\sigma}\,\partial_{\mu} e^{\sigma}{}_{a}\,.
\end{equation}
This connection 1-form has an associated \emph{exterior covariant derivative}, that acts on forms $\dfal_{a...}{}^{b...}$ as
  \begin{equation}
    \Dex\dfal_{a...}{}^{b...}\ =\ \dex\dfal_{a...}{}^{b...} \ + \ \dfom_{c}{}^{b}\wedge\dfal_{a...}{}^{c...}\ + \ ...\ -\ \dfom_{a}{}^{c}\wedge\dfal_{c...}{}^{b...}\ -  \ ...\,.
    \label{extcovder}
  \end{equation}
The curvature and torsion 2-forms and the non-metricity 1-form are then defined as 
  \begin{align}
    \dfR_{a}{}^{b} \, &  = \ \dex\dfom_{a}{}^{b}\ + \ \dfom_{c}{}^{b}\wedge\dfom_{a}{}^{c}\,,\\
    \dfT^{a}      \,  &  =\ \dex\cofr^{a}+\dfom_{c}{}^{a}\wedge\cofr^{c} \quad = \Dex\cofr^{a} \,,\\ 
    \dfQ_{ab}      \, &  =\ -\Dex g_{ab}\,.
  \end{align}
whose components are indeed those of the corresponding curvature, torsion and non-metricity tensors we introduced previously:
  \begin{align}
    \dfR_{a}{}^{b} \, & = \ e^\rho{}_a \, e^b{}_\lambda \, \big(\tfrac{1}{2}\, \cR_{\mu\nu \rho}{}^{\lambda}\, \dex x^{\mu}\wedge\dex x^{\nu}\big)\,,\\
    \dfT^{a}       \, & =\  (\partial_{[\mu} e^a{}_{\nu]} + \omega_{[\mu|c}{}^a e^c{}_{|\nu]}) \, \dex x^{\mu}\wedge\dex x^{\nu}
\ = \ e^a{}_\lambda\,  \big(\tfrac{1}{2}\, T_{\mu\nu}{}^{\lambda}\, \dex x^{\mu}\wedge\dex x^{\nu}\big)
    \,,\\
    \dfQ_{ab}      \, & = \ e^\nu{}_a \, e^\rho{}_b \, (Q_{\mu\nu\rho} \, \dex x^{\mu}) \,.
  \end{align}

In the presence of a metric, the indices of the connection 1-form can be freely raised  and lowered. In addition, it is not difficult to prove that if we extract from $\dfom_{ab}$ its metric-compatible part $\bar \dfom_{ab}$, the rest turns out to be a symmetric tensor proportional to the non-metricity,
\be
\dfom_{ab} \ = \ \bar \dfom_{ab} \ + \ \half \, \dfQ_{ab}\,.
\label{splitting}\ee
It is important to remark that $\bar \dfom_{ab}$ is a connection in its own right (metric-compatible by definition), whose torsion depends on the torsion and the non-metricity of $\dfom_a{}^b$. Indeed, $\bar\dfom_{ab}=\dfom_{[ab]}+\half \dex g_{ab}$ and, thanks to the $\mathrm{GL}(\dimM,\mathbb{R})$ gauge choice we are assuming throughout this paper, $\dex g_{ab}=0$, the equation \eqref{splitting} can be seen simply as the decomposition of $\dfom_{ab}$ into its antisymmetric and symmetric parts.

Taking into account the projective invariance (\ref{projsymm}) of the Einstein-Palatini action, it is also useful to split the non-metricity $\dfQ_{ab} = \dfQTrLess{}_{ab} + \half \dfQ_c{}^c g_{ab}$ into its trace $\dfQ_c{}^c$ and traceless degrees of freedom $\dfQTrLess{}_{ab}$ (see Appendix \ref{IrrepsTQ}). Therefore, the general connection decomposes as
\be
\dfom_{ab} \ = \ \bar \dfom_{ab} \ + \ \half \, \dfQTrLess{}_{ab} \ + \ \tfrac{1}{4}\,\dfQ_c{}^c g_{ab}.
\label{omegasplitting}
\ee
Each of these three fields ($\bar \dfom_a{}^b$, $\dfQTrLess_{ab}$ and $\dfQ_c{}^c$) can be treated as an independent field, giving a system of three equations of motion, equivalent to the equation of motion of $\dfom_a{}^b$. 

The curvature 2-form of  $\dfom_a{}^b$ and $\bar \dfom_a{}^b$ are hence related as
\be
\dfR_a{}^b(\dfom) \ =  \ \bar \dfR_a{}^b(\bar \dfom) \ + \ \tfrac{1}{2}\bar\Dex \dfQTrLess_a{}^b
\ + \ \tfrac{1}{4} \dex{\dfQ}_c{}^c \, \delta_a{}^b
\ - \  \tfrac{1}{4} \dfQTrLess_a{}^c\wedge  \dfQTrLess_c{}^b,
\ee
where $\bar\Dex$ is the (metric-compatible) exterior covariant derivative associated with  $\bar \dfom_a{}^b$. This allows us to express the two-dimensional Einstein-Palatini action  (\ref{dim=2EH}) as
\bea
S_1^{(2)} \ = \ \frac{1}{2\kappa} \int \LCtensor^a{}_b \, \dfR_a{}^b(\dfom)
\ =  \ \frac{1}{2\kappa} \int  \LCtensor_{ab} \, \Bigl[\bar \dfR{}^{ab}(\bar \dfom) \ - \  \tfrac{1}{4}\, \check\dfQ{}^{ac}\wedge \check\dfQ_c{}^b\Bigr],
\label{dim=2EH-forms}
\eea
where we have introduced the two-dimensional Levi-Civita tensor $ \LCtensor_{a_1 a_2}  = \sqrt{|\det(g_{ab})|} \,\LCsymbol_{a_1 a_2}$, canonically associated to the metric (see Appendix \ref{AppendixDiffForms} for the general definition). The terms $\bar\Dex \check \dfQ{}^{ab}$ and $\dex\dfQ_c{}^c g^{ab}$ drop out of the action, due to the antisymmetry of $\LCtensor_{ab}$. Note that $\dfQ_c{}^c$ does not appear in the Einstein-Palatini action, and hence remains undetermined, in agreement with the projective symmetry. As we will see, this property also holds for the general Lovelock Lagrangians (critical or non-critical)  \cite{JJO}. 

Considering an orthonormal gauge $g_{ab}=\eta_{ab}$, we find that
\begin{equation}
\LCtensor_{ab} \, \bar \dfR{}^{ab}(\bar\dfom) \ = \ \LCtensor_{ab} \, \dex\bar\dfom^{ab} \ = \  \dex (\LCtensor_{ab}\,\bar\dfom^{ab})\,,
\end{equation}
where in the first step we have used that $\LCtensor_{ab} \ \bar\dfom^{ac} \wedge \bar\dfom_c{}^b =  0$, due to the antisymmetry of both $\LCtensor_{ab}$ and $\bar\dfom^{ab}$ and the fact that the theory lives in $\dimM=2$ (the indices $a$, $b$ and $c$ have to be all different, but at the same time can only take values in the set $\{1,2\}$), while in the second step we used the fact that $\dex\LCtensor_{ab}=\dex\LCsymbol_{ab}=0$. The two-dimensional Einstein-Palatini action therefore reduces to
\begin{equation}
  S_1^{(2)} \ = \ \frac{1}{2\kappa} \int \Bigl[  \dex(\LCtensor_{ab} \, \bar\dfom^{ab}) \ - \ \tfrac{1}{4}\, \LCtensor_{ab} \, \dfQTrLess^{ac}\wedge \dfQTrLess_c{}^b\Bigr]\,.
\label{dim=2EH-forms-final}
\end{equation}
So we find that the two-dimensional metric-affine Einstein term cannot be written as a total derivative, unless the connection verifies $\dfQTrLess_{ab}=0$. Let us insist that $\bar\dfom_a{}^b$ is an arbitrary metric-compatible connection (that might include torsion). Indeed it is only the (traceless part of) the non-metricity what spoils the topological character of the theory.

Finally, let us quickly derive the results of the previous subsection in the language of differential forms. As $\dfQTrLess{}_{ab}$ is the only dynamical variable of   (\ref{dim=2EH-forms-final}), its equation of motion can easily be calculated as
\begin{equation}
 0 = \delta_{\dfQTrLess} S^{(2)}_1 \ = \ \frac{1}{2\kappa} \int \delta \dfQTrLess{}^{ac}\wedge \big(- \tfrac{1}{2} \LCtensor_{ab} \,\dfQTrLess_c{}^b\big)
  \quad  \Rightarrow \quad
 \LCtensor_{ab} \,\dfQTrLess_c{}^{b} \ = \ 0\,,
\end{equation}
whose only solution is the one we found in (\ref{EH-Q}),
\begin{equation} \dfQTrLess_{ab}=0\,. \end{equation}

%%%%%%%%%%%%%%%%%%%%%%%%%%%%%%%%%%%%%%%%%%%%%%%%%%%%%%%%%
\section{The Gauss-Bonnet-Palatini action in \texorpdfstring{$\dimM=4$}{n=4}}
\label{dim=4GB}

Our next step will be to look at the second order Lovelock term in the corresponding critical dimension, the four-dimensional Gauss-Bonnet-Palatini theory, whose action is given by\footnote{Recall that this action can also be written in components as 
\be
\dfL_{2}^{(4)}     \ = \ \left[\cR^2 -\cR_{\mu\nu} \cR^{\nu\mu} + 2 \cR_{\mu\nu} \cR^{(2)\nu\mu} -\cR^{(2)}{}_{\mu\nu} \cR^{(2)\nu\mu} + \cR_{\mu\nu\rho\lambda} \cR^{\rho\lambda\mu\nu}\right] \sqrt{|g|}\ \dex^4x\,,
\nonumber
 \ee
where $\cR^{(2)}{}_\mu{}^\nu = g^{\rho\sigma} \cR_{\mu\rho\sigma}{}^\nu$ is the second contraction of the Riemann tensor.}
\be
\dfL_{2}^{(4)} \ = \ \LCtensor^a{}_b{}^c{}_d \,\dfR_{a}{}^{b}(\dfom)\wedge\dfR_{c}{}^{d}(\dfom).
\label{LagGBPCrit}
\ee
Unfortunately, the dynamics of this case is already too complicated to solve the theory completely, as we have done for the $\dimM = 2$ Einstein-Palatini theory. However we will argue that also here it is the traceless part of the non-metricity $\dfQTrLess_{ab}$ what prevents the theory from being a boundary term, first by writing the action (\ref{LagGBPCrit}) as a total derivative plus $\dfQTrLess$-dependent terms and then presenting some specific field configurations that do not obey the equations of motion.

%%%%%%%%%%%%%%
\subsection{The \texorpdfstring{$\dimM=4$}{n=4} Gauss-Bonnet-Palatini term is not a total derivative}

In order to isolate a total derivative term in the action (\ref{LagGBPCrit}), it is again useful to split the connection 1-form into its antisymmetric, traceless symmetric and trace parts, in a way similar to (\ref{omegasplitting}). In $\dimM = 4$ we have
\be
\dfom_{ab} \ = \ \bar \dfom_{ab} \ + \ \half \, \dfQTrLess{}_{ab} \ + \ \tfrac{1}{8}\,\dfQ_c{}^c g_{ab},
\label{omegasplitting2}
\ee
yielding the Riemann tensor to split as
\be
\dfR_a{}^b(\dfom) \ =  \ \bar \dfR_a{}^b(\bar \dfom) \ + \ \tfrac{1}{2}\bar\Dex \dfQTrLess_a{}^b
\ + \ \tfrac{1}{8} \dex{\dfQ}_c{}^c \, \delta_a{}^b
\ - \  \tfrac{1}{4} \dfQTrLess_a{}^c\wedge  \dfQTrLess_c{}^b. 
\ee
Substituting this into (\ref{LagGBPCrit}), the action becomes
\bea
\dfL_{2}^{(4)} \ = \ \LCtensor_{abcd}{}\, \Bigl[ \bar\dfR{}^{ab}\wedge\bar\dfR{}^{cd}
  \ - \ \half \bar\dfR{}^{ab}\wedge\dfQTrLess{}^{cf}\wedge\dfQTrLess_f{}^{d}
  \ + \ \tfrac{1}{16} \dfQTrLess{}^{ae} \wedge\dfQTrLess_e{}^b \wedge \dfQTrLess{}^{cf} \wedge \dfQTrLess_f{}^d\Bigr].
\eea
The first term is the four-dimensional Euler characteristic and can easily be written as a total derivative (see for example \cite{YP, HMMN}). Choosing again an orthonormal gauge ($g_{ab}=\eta_{ab}$), we therefore find that the action (\ref{LagGBPCrit}) is of the form
\be
\dfL_{2}^{(4)} \ = \ \dex \dform{{\cal C}} \ - \ \LCtensor_{abcd}{}\, 
  \Bigl[\half\bar\dfR{}^{ab} \wedge \dfQTrLess{}^{cf} \wedge \dfQTrLess_f{}^{d}
  \ - \ \tfrac{1}{16} \dfQTrLess{}^{ae} \wedge\dfQTrLess_e{}^b \wedge \dfQTrLess{}^{cf} \wedge \dfQTrLess_f{}^d\Bigr], 
\ee
where
\begin{equation}
\dform{{\cal C}} = \LCtensor^a{}_b{}^c{}_d \,\Bigl[\, \bar\dfR_a{}^b \wedge \bar \dfom_c{}^d
     \ + \ \tfrac{1}{3}  \bar\dfom_a{}^b \wedge \bar\dfom_c{}^f \wedge \bar\dfom_f{}^d \Bigr]\,.
\end{equation}
Again we find that the $\dimM = 4$ Gauss-Bonnet-Palatini term can only be written as a total derivative for affine connections with $\dfQTrLess{}_{ab} = 0$.

%%%%%%%%%%%%%%%%%%%
\subsection{The equations of motion of \texorpdfstring{$\dimM=4$}{n=4} Gauss-Bonnet theory}
\label{subsection:EoMCriticalGBP}

Let us now look at the dynamics of this theory. The equation of motion of the coframe $\cofr^{a}$ is the easiest to compute for the action in the form \eqref{LagGBPCrit}. As explained in Appendix \ref{AppendixEoM}, the equation of motion for a general Lagrangian of the form $\dfL(g_{ab},\cofr^a,\dfR_{a}{}^{b}(\dfom))$ is given by\footnote{Here we have introduced the interior product $\dint{}$. See Appendix \ref{AppendixDiffForms} for the definition and examples.}
\be
0\ = \ \dint{\boldsymbol{e}_{m}} \dfL
\ - \ (\dint{\boldsymbol{e}_{m}}\dfR_{p}{}^{q}) \wedge \left(\frac{\partial\dfL}{\partial\dfR_{p}{}^{q}} \right). 
\ee
Specifically, for the Gauss-Bonnet term \eqref{LagGBPCrit}, we find that
\bea
0 &=&  \dint{\boldsymbol{e}_{m}} \Bigl[\LCtensor^a{}_b{}^c{}_d \,\dfR_{a}{}^{b}\wedge\dfR_{c}{}^{d} \Bigr]
\ - \ (\dint{\boldsymbol{e}_{m}}\dfR_{p}{}^{q}) \wedge \Bigl[2\, \LCtensor^a{}_b{}^p{}_q \,\dfR_{a}{}^{b}\Bigr] \nnw
&=& 2\, \LCtensor^a{}_b{}^c{}_d\, (\dint{\boldsymbol{e}_{m}}\dfR_{a}{}^{b})\wedge\dfR_{c}{}^{d}
\ - \ 2\, \LCtensor^a{}_b{}^p{}_q \,  (\dint{\boldsymbol{e}_{m}}\dfR_{p}{}^{q}) \wedge \dfR_{a}{}^{b} \nnw
&=& 0. \label{GBcoframeeqn}
\eea
In other words, the equation of motion of the coframe is automatically satisfied, for any frame $\boldsymbol{e}_{a}$ and connection configuration $\dfom_a{}^b$. This property is not surprising for metric-compatible connections, as we have just proven that in that case the Lagrangian is a total derivative. However, for general connections with $\dfQTrLess{}_{ab} \neq 0$, this is much less obvious.  We will see in the next section that this property is true for all critical metric-affine Lovelock terms.

As we explained above, the equation of motion of the connection is equivalent to the system of dynamical equations for the fields $\dfQTrLess_{ab}$ and $\bar\dfom_a{}^b$, given respectively by
\bea
0 & =& \LCtensor_{abcd} \, \left[\bar{\dfR}{}^{ab}\, - \, \tfrac{1}{4}\dfQTrLess_{f}{}^{a} \wedge \dfQTrLess{}^{bf}\right] \wedge \dfQTrLess{}^{dm} \,, \\ [.2cm]
0 & =& \bar{\Dex} \left[\dfQTrLess_{c}{}^{a}\wedge\dfQTrLess{}^{bc}\right] \label{eomQTrless-GB}\,.
\eea
The last equation has been contracted by another Levi-Civita tensor to eliminate the one coming from the Lagrangian, using \eqref{contractionLCtensor}. In principle, this produces the antisymmetrisation in $\{ab\}$, which can be dropped, since, $\dfQTrLess_{ab}$ being a 1-form, the combination $\dfQTrLess_{c}{}^{a}\wedge\dfQTrLess{}^{bc}$ is already antisymmetric.

As we mentioned earlier, these equations of motion are too complicated to solve in their full generality. However to illustrate the non-topological nature of the action  (\ref{LagGBPCrit}), it is sufficient to come up with a field configuration that does not satisfy the equations (\ref{eomQTrless-GB}), as this would proof that the equations of motion do impose some non-trivial conditions. 

For instance consider the gravitational field configuration
\bea
&& g_{ab}  \ = \ \eta_{ab}\,, \hsp{2cm}
\bar \dfom^{ab} \ = \ \mathring{\dfom}^{ab} \, + \, f \dfal^{[a}\delta_{t}^{b]} \,,
\nonumber\\ [.2cm]
&&  \cofr^{a} \ = \ \dex x^{a}\,,  \hsp{2cm}
  \dfQTrLess{}^{ab} \ = \ 2 \dfal^{(a}\delta_{t}^{b)}\,.\label{ansatzGB}
\eea
where $\eta_{ab}$ is the Minkowski metric, $f$ is an arbitrary function and
\begin{equation}
  \dfal^a \ = \ \eN^t \, \left(\delta_{y}^{a}\,\dex y\ +\ \delta_{z}^{a}\,\dex z\right)\,.
\end{equation}
Let us remark a couple of details. First, note that this Ansatz is consistent with the fact that $\dfQTrLess{}^{ab}$ is traceless, since $\dfal_c\delta_{t}^c=0$. Furthermore, observe also that we can everywhere drop the Levi-Civita connection, since the associated metric is Minkowski and the latin indices are referred to the Cartesian basis of the space, as can be seen in the expression for  $\cofr^{a}$ in \eqref{ansatzGB}. 

Under these conditions we have that
\begin{equation}
  \dfQTrLess_{c}{}^{a}\wedge\dfQTrLess{}^{bc} = \dfal^{a} \wedge \dfal^{b} \,,
\end{equation}
and, with this in mind, it is not difficult to check that the Ansatz is a counterexample that violates the condition (\ref{eomQTrless-GB}):
\begin{equation}
   \bar{\Dex} \left[\dfQTrLess_{c}{}^{a}\wedge\dfQTrLess{}^{bc}\right] 
     \ = \ \dex\left[\dfal^{a} \wedge \dfal^{b}\right]
     \ = \ 2\eN^{2t} \left(\delta_y^a \delta_z^b \, -\, \delta_y^b \delta_z^a \right)\dex t \wedge\dex y \wedge\dex z  \ \neq \ 0\,.
\end{equation}
It is worth remarking that this inequality holds in the entire manifold, since this set of coordinates is globally defined. This result proves that the metric-affine Gauss-Bonnet term in $\dimM=4$ is not a trivial theory (in the sense that it cannot be written as a total derivative), since only some configurations of the fields are allowed by the equations of motion.

%%%%%%%%%%%%%%%%%%%%%%%%%%%%%%%%%%%%%%%%%%%%%%%%%%%%%%%%%
\section{The \texorpdfstring{$k$}{k}-th order metric-affine Lovelock term in \texorpdfstring{$\dimM=2k$}{n=2k}}
\label{dim=2kLovelock}

\subsection{Proving that the theory is not a boundary term}

Finally we will analyse the general case of the $k$-th order Lovelock term in  $\dimM=2k$ dimensions, defined as 
\begin{equation}
  \dfL_{k}^{(2k)} \ = \ \LCtensor^{a_1}{}_{a_2} ...{}^{a_{2k-1}}{}_{a_{2k}}\dfR_{a_1}{}^{a_2} \wedge... \wedge \dfR_{a_{2k-1}}{}^{a_{2k}} \,.
  \label{n=2kaction}
\end{equation}
Again, splitting the connection $\dfom_{ab}$ as
\be
\dfom_{ab} \ = \ \bar \dfom_{ab} \ + \ \half \, \dfQTrLess{}_{ab} \ + \ \tfrac{1}{2\dimM}\,\dfQ_c{}^c g_{ab},
\label{omegasplitting3}
\ee
it is straightforward to check that the action can be written as a power series in $\bar \dfR$ and $\dfQTrLess \wedge \dfQTrLess$ terms, 
\bea
\dfL_{k}^{(2k)} &=& \LCtensor_{a_1 \dots a_{2k}}\ \sum_{m=0}^k \,
\frac{1}{4^{k-m}} \ \frac{k!}{m! (k-m)!} \, 
 \  \bar \dfR{}^{a_1a_2} \wedge\, ... \, \wedge  \bar\dfR{}^{a_{2m-1}a_{2m}}\wedge  \nnw
 && \hsp{2cm}
 \wedge \,\dfQTrLess{}^{a_{2m+1}f_1} \wedge \dfQTrLess_{f_1}{}^{a_{2m+2}} \wedge \, ... \, \wedge\dfQTrLess{}^{a_{2k-1}f_{k-m}} \wedge \dfQTrLess_{f_{k-m}}{}^{a_{2k}} \,, \label{GBLagSplitting}
 \eea
 of which only  the $m=k$ term $\bar \dfR_{a_1}{}^{a_2} \wedge\, ... \, \wedge  \bar\dfR_{a_{2k-1}}{}^{a_{2k}}$ is in fact a boundary term \cite{TZ, Zanelli} (see also \cite{DMO, CPRS}). This can be easily seen using the Bianchi identity of $\bar \Dex$ to show that it is a closed form and, hence, locally exact by Poincar{\'e} lemma. For this reason, we will eliminate this term for the rest of this section, as it does not contribute to the equations of motion.

As we did in the Gauss-Bonnet case, we will construct a field configuration that does not satisfy the equation of motion of $\bar\dfom_a{}^b$, proving that the latter is not automatically satisfied. The equation of motion of $\bar\dfom_a{}^b$ can be found in \eqref{GeneralEoMbarConn}, and the right hand side of the equation is given by\footnote{The hat indicates that the action is considered with respect to $\bar\dfom_a{}^b$ and $\dfQTrLess_{ab}$, i.e.  
\[
\hat{S}_k^{(2k)}[g,\cofr,\bar\dfom,\dfQTrLess] =  S_k^{(2k)}[g,\cofr,\dfom(\bar\dfom,\dfQTrLess)]\,.
\]
}
\begin{align}
g_{ca}\frac{\delta \hat{S}_k^{(2k)}}{\delta \bar\dfom_{c}{}^{b}} 
 & \, = \ \LCtensor_{ab a_3 \dots a_{2k}}\ \sum_{m=1}^{k-1} \,
\frac{1}{4^{k-m}} \ \frac{k!}{m! (k-m)!} \, 
 \  \bar \dfR{}^{a_3a_4} \wedge\, ... \, \wedge  \bar\dfR{}^{a_{2m-1}a_{2m}}\wedge \nonumber \\ 
 & \qquad \qquad 
 \wedge \, \bar \Dex \Big[\dfQTrLess{}^{a_{2m+1}f_1} \wedge \dfQTrLess_{f_1}{}^{a_{2m+2}} \wedge \, ... \, \wedge\dfQTrLess{}^{a_{2k-1}f_{k-m}} \wedge \dfQTrLess_{f_{k-m}}{}^{a_{2k}} \Big] \,, \label{GeneralLovelockEoMbarOm}
\end{align}
where we have taken into account the Bianchi identity $\bar \Dex \bar\dfR_a{}^b=0$.

Consider now the  Ansatz
\bea
&& g_{ab}  \ =\ \eta_{ab}\,, \hsp{2cm}
\bar \dfom^{ab} \ = \ \mathring{\dfom}^{ab} \,,\\ [.2cm]
&& \cofr^{a} \ = \ \dex x^{a}\,, \hsp{2cm}  
\dfQTrLess{}^{ab} \ =  \ 2 \dfal^{(a}\delta_{t}^{b)}\,,
\eea
where we now define
\begin{equation}
  \dfal^a \ = \ \eN^t\, \left(\delta_{3}^{a}\dex x^3 \, + \, ... \, + \, \delta_{2k}^{a}\dex x^{2k} \right) \,,
\end{equation}
and the $x^a$ take values in the set $\{x^1=t,~x^2,...,~x^{2k}\}$. Note that this Ansatz is consistent with $\dfQTrLess_c{}^c=0$. Since our connection $\bar \dfom^{ab}$ is flat, we have that $\mathring\dfR{}^{ab}=0$ and only the  $m=1$ term in \eqref{GeneralLovelockEoMbarOm} survives, as it is the only one that does not contain $\bar \dfR_{ab}$. For the same reason, the covariant exterior derivative reduces to the ordinary exterior derivative: $\bar\Dex=\dex$. The equation \eqref{GeneralLovelockEoMbarOm} then simplifies to
\begin{align}
\eta_{ca}\frac{\delta \hat{S}_k^{(2k)}}{\delta \bar\dfom_{c}{}^{b}} 
& \, = \ \LCtensor_{ab a_3 \dots a_{2k}}\ \,
\frac{k}{4^{k-1}} \ \dex\Big[\dfQTrLess{}^{a_{3}f_1} \wedge \dfQTrLess_{f_1}{}^{a_{4}} \wedge \, ... \, \wedge\dfQTrLess{}^{a_{2k-1}f_{k-1}} \wedge \dfQTrLess_{f_{k-1}}{}^{a_{2k}} \Big] \,.
\label{omegaeqn}
\end{align}

As in the Gauss-Bonnet case, we have that $\dfQTrLess_{c}{}^{a}\wedge\dfQTrLess{}^{bc} = \dfal^{a} \wedge \dfal^{b}$, so the equation (\ref{omegaeqn}) can be rewritten as
\bea
 && \frac{-4^{k-1}}{k 2!(2k-2)!} \ \LCtensor_c{}^{b a_3 \dots a_{2k}}\ \,
 \frac{\delta \hat{S}_k^{(2k)}}{\delta \bar\dfom_{c}{}^{b}} \nonumber \\[.2cm]
 && \qquad\qquad =  \ \dex \big(\dfal^{a_3} \wedge \, ... \, \wedge \dfal^{a_{2k}}\big)
 \nonumber \\ [.2cm]
 &&  \qquad\qquad = \ 2(k-1) \, \eN^{2(k-1)t} \ (2k-2)! \, \delta_3^{[a_3} ... \delta_{2k}^{a_{2k}]} \ \dex t \wedge\dex x^3 \wedge\dex x^4\wedge...\wedge \dex x^{2k}\,.
\eea
Again, it is easy to see that this expression is non-zero in the entire manifold, except for
$k=1$. However, as we already solved the $k=1$ case completely in Section \ref{EH}, in practice we are only interested in $k>1$. In summary, by finding a field configuration that does not satisfy the  $\bar\dfom_a{}^b$ equation, we have extended the argument from the Gauss-Bonnet case to general $k$-th order critical Lovelock theories, proving that in general the equations of motion impose non-trivial conditions.

%%%%%%%%%%%
\subsection{Exploring non-trivial solutions of the critical case of arbitrary \texorpdfstring{$k$}{k}}

According to the previous result, it makes sense to search for non-trivial solutions for the critical metric-affine Lovelock theory of arbitrary order.

It is not difficult to see that the triviality of the equation of the coframe, proven in Section \ref {dim=4GB} for the Gauss-Bonnet case, is in fact a general property for all critical Lovelock theories. Indeed, it is straightforward to generalise the argument given in (\ref{GBcoframeeqn}) to arbitrary $k$. Yet, there is also a particularly simple way of seeing this, looking at the direct variation with respect to the coframe given in (\ref{GeneralEoMCofr}),
\begin{equation}
0\ = \ \frac{\partial \dfL}{\partial \cofr^a}\,.
\label{EoMCofrGen1}
\end{equation}
In contrast to the (non-critical) $k$-th order Lovelock term in arbitrary dimensions $\dimM$,\footnote{Here we have introduced the Hodge dual $\star$ associated to the metric structure. See Appendix \ref{AppendixDiffForms} for its explicit expression.}
\begin{align}
  \dfL_{k}^{(\dimM)} 
& = \ \dfR^{a_1 a_2}\wedge...\wedge\dfR^{a_{2k-1} a_{2k}}\wedge \star (\cofr_{a_1}\wedge...\wedge \cofr_{a_{2k}}) \\ 
& = \ \frac{1}{(\dimM-2k)!} \ \LCtensor_{a_1...a_{2k} b_1...b_{\dimM-2k}} \, \dfR^{a_1 a_2}\wedge...\wedge\dfR^{a_{2k-1} a_{2k}}\wedge \cofr^{b_1}\wedge...\wedge \cofr^{b_{\dimM-2k}} \,,
\end{align}
the critical $k$-th order Lovelock term (\ref{n=2kaction}) has no explicit dependence on the coframe, implying that the equation (\ref{EoMCofrGen1}) is trivially satisfied. A similar proof for critical Lovelock terms for arbitrary order in the metric formalism can be found in \cite{CM}.

Consequently the only remaining equation is the one for the connection. Varying the action with respect to $\dfom_a{}^b$, we find  (see \eqref{GeneralEoMConn})
\begin{align}
  0 & = \Dex\LCtensor^{a_1}{}_{a_2}...{}^{a}{}_{b}\wedge\dfR_{a_1}{}^{a_2}\wedge...\wedge\dfR_{a_{2k-3}}{}^{a_{2k-2}}\\
    & = \Big[ \delta^d_{a_{1}} \LCtensor_{ca_2 ...a_{2k-2}ab} \ + \ ... \ + \ \delta^d_{a_{2k-3}} \LCtensor_{a_{1}...a_{2k-4}ca_{2k-2}ab} \nonumber\\
    & \qquad\qquad + \delta^d_a \, \LCtensor_{a_1 ...a_{2k-2}cb}\Big] \dfQTrLess{}^c{}_d \wedge \dfR^{a_1 a_2}\wedge...\wedge\dfR^{a_{2k-3}a_{2k-2}}\,. \label{EoMConnGeneral}
\end{align}

We will consider an arbitrary coframe $\cofr^a$, such that $g_{ab}$ are constant, as this is an hypothesis we have been using from the beginning. By observing the equation \eqref{EoMConnGeneral} one can easily deduce a series of non-trivial connections that, together with that coframe, constitute solutions of the theory:
\begin{itemize}
  \item \textbf{Solutions for arbitrary $k$:} In general, the equation is fulfilled by any connection with identically zero traceless part $\dfQTrLess_{ab}$ (i.e. $Q_{\mu\nu\rho}=V_{\mu} g_{\nu\rho}$ for some 1-form $V_{\mu}$). Note that an interesting subcase is the connection \eqref{GBPsol}, that was presented in \cite{JJO} as a particular non-trivial solution for the $k=2$ case, but conjectured to hold for arbitrary $k$.

  \item \textbf{Solutions for $k>1$:} For the second or higher order Lovelock critical Lagrangian, in the equation \eqref{EoMConnGeneral} there is at least one curvature as a global factor, so any teleparallel connection ($\dfR_{c}{}^{d} =0$) is a solution. Indeed, we can infer a slightly more general result: any connection satisfying
  \begin{equation}
    \dfQTrLess_{ab}\wedge \dfR_{c}{}^{d} = 0
  \end{equation}
  gives a solution of the equations of motion.

  \item \textbf{Solutions for $k>2$:} In these cases, there are two or more curvatures in the equation of motion. So any connection such that $\dfR_{ab} = \dfal_{ab}\wedge \dfk$ for certain 1-forms $\dfal_{ab}$ and $\dfk$, is a solution. In this category we find for example those studied in the context of gravitational waves in Poincar\'e gravity \cite{Obuk}, where $\dfk$ is the dual form of the wave vector.
\end{itemize}

%%%%%%%%%%%%%%%%%%%%%%%%%%%%%%%%%%%%%%%%%%%%%%%%%%%%%%%%%
\section{Conclusions}
\label{Conclusions}

In this paper we studied metric-affine Lovelock theories in critical dimensions, i.e. the $k$-th order term in $\dimM = 2k$ dimensions. Where it is standard lore that critical Lovelock terms are topological invariants, when equipped with the Levi-Civita connection, we have proven that this is not the case for general affine connections. In particular, it is the traceless part of the non-metricity $\dfQTrLess{}_{ab} = \dfQ_{ab} - \tfrac{1}{n} \dfQ_{c}{}^c g_{ab}$ that adds extra dynamical degrees of freedom to the Lagrangian.

We have performed a case by case study, starting with the lowest order case, the Einstein-Palatini action, and gradually moving up till the general $k$-th order term. For the two-dimensional Einstein-Palatini case, we have found the most general solution. It is given by an arbitrary (constant) metric $g_{ab}$, an arbitrary coframe $\cofr^a$ and a connection that is restricted to have vanishing  $\dfQTrLess_{ab}$. This constraint, which affects four of the eight degrees of freedom of $\dfom_a{}^b$, is an indicator that the Lagrangian cannot be an exact form, as it imposes non-trivial conditions on the field configurations. Indeed, by decomposing the connection into its metric-compatible part $\bar \dfom_a{}^b$ and its non-metricity $\dfQ_{ab}$ and rearranging the terms in the Lagrangian, we have shown explicitly that the Einstein-Palatini term takes the form of a boundary term that depends on $\bar \dfom_a{}^b$ plus a non-exact form that depends on  $\dfQTrLess_{ab}$.

The analysis we made for the Gauss-Bonnet-Palatini case is in fact a particular example of the general critical metric-affine Lovelock Lagrangian, so we will discuss all $k>1$ cases together. In all of them, the theory is too complicated to solve completely. However, we were able to proof the dynamical nature of the theory by providing a counterexample that violates at least one of the equations of motion. This implies that the Lagrangian cannot be reduced to a boundary term, since in that case the equations of motion would have been identically satisfied for all configurations. Again we find in all the cases that the Lagrangian can be written as an exact form plus $\dfQTrLess_{ab}$-dependent terms. Therefore, in the metric-affine formulation, it is not possible to rewrite curvature invariants in terms of other ones through integration by parts, since additional terms depending on the traceless part of the non-metricity come into play.

Then, we showed that the coframe equation is always identically satisfied, but that this is not the case for the connection equation, as our counterexamples illustrate. Finally we suggested some non-trivial families of solutions for different values of $k$. 

As we mentioned earlier, the traceless part of the non-metricity $\dfQTrLess_{ab}$ appears as the main agent that prevents the theory from being a boundary term. It is worth remarking that $\dfQTrLess_{ab}$ is not an irreducible part of the non-metricity under the $\mathrm{GL}(\dimM,\mathbb{R})$ group, as can be seen in the Appendix \ref{IrrepsTQ}. It would be interesting to split $\dfQTrLess_{ab}$ into its three irreducible parts and see whether they are all dynamical or whether the non-topological nature of the critical Lovelock actions comes only from a specific part.

So far, we have only looked at the dynamics of separate Lovelock terms in critical dimensions. It would be interesting to investigate the full Lovelock theory including all the terms with $k \leq \lfloor \dimM/2 \rfloor $, and look for non-trivial solutions of the full theory. We leave this for future research.

%%%%%%%%%%%%%%%%%%%%%%%%%%%%%%%%%%%%%%%%%%%%%%%%%%%%%%%%
%%%%%%%%%%%%%%%%%%%%%%%%%%%%%%%%%%%%%%%%%%%%%%%%%%%%%%%%
%%%%%%%%%%%%%%%%%%%%%%%%%%%%%%%%%%%%%%%%%%%%%%%%%%%%%%%%
%\newpage
\vspace{.6cm}
\noindent
{\bf Acknowledgements}\\
The authors would like to thank Jos\'e Beltr\'an Jim\'enez, Tomi Koivisto, Yu Nakayama,  Jos\'e Alberto Orejuela, Miguel S\'anchez and Jorge Zanelli for useful discussions
and Konstantinos Pallikaris for detecting some typos in the previous version. This work was partially supported by the Spanish Ministry of Economy and Competitiveness (FIS2016-78198-P), the Junta de Andaluc\'ia (FQM101 and Proyecto SOMM17/6104/UGR)
and the Unidad de Excelencia UCE-PP2016-02 of the Universidad de Granada. A.J.C. is supported by a PhD contract of the program FPU 2015 with reference FPU15/02864 (Spanish Ministry of Economy and Competitiveness).

%\newpage
%%%%%%%%%%%%%%%%%%%%%%%%%%%%%%%%%%%%%%%%%%%%%%%%%%%%%%%
\appendix
%%%%%%%%%%%%%%%%%%%%%%%%
\section{Irreducible parts of torsion and non-metricity}
\label{IrrepsTQ} 

We will give a quick review of the decomposition of the torsion and the non-metricity in their irreducible parts. A more detailed discussion in terms of differential forms can be found in \cite{McCrea, HMMN}.  

In $\dimM$ dimensions, the torsion can in general be divided in three irreducible parts, 
\begin{equation}
T_\mn{}^\rho \ = \  T^{\rm (tr)}{}_\mn{}^\rho \ + \ T^{\rm (a)}{}_\mn{}^\rho \ + \ T^{\rm (tn)}{}_\mn{}^\rho,
\end{equation}
where
\bea
T^{\rm (tr)}{}_\mn{}^\rho & =& \frac{2}{\dimM-1} \, T_{[ \mu |\sigma|}{}^\sigma \, \delta_{\nu]}^\rho \,, \\[.2cm]
T^{\rm (a)}{}_\mn{}^\rho  &=& g^{\rho\sigma} \, T_{[\mn\sigma]}\,,\\[.2cm]
T^{\rm (tn)}{}_\mn{}^\rho &=& T_\mn{}^\rho \, - \, T^{\rm (tr)}{}_\mn{}^\rho \, - \, T^{\rm (a)}{}_\mn{}^\rho \,.
\eea
which are respectively the trace, the completely antisymmetric part and  the remaining trace-free part. In particular  $T^{\rm (tn)}{}_\mn{}^\rho$ satisfies the following cyclic property:
\begin{equation}
T^{\rm (tn)}{}_\mnr \ + \  T^{\rm (tn)}{}_{\rho\mu\nu} \ + \ T^{\rm (tn)}{}_{\nu\rho\mu} \ = \ 0.
\end{equation}
Note that in general the $ \half \dimM^2 (\dimM-1)$ components of the torsion are distributed as follows over the three irreducible parts: $T^{\rm (tr)}{}_\mn{}^\rho$ has $\dimM$ independent components, $T^{\rm (a)}{}_\mn{}^\rho$ has $\tfrac{1}{6}\dimM (\dimM-1)(\dimM-2)$ and $T^{\rm (tn)}{}_\mn{}^\rho$ the remaining $\frac{1}{3}\dimM(\dimM^2-4)$.  

The decomposition of the non-metricity is a bit more involved. In this case, there are four irreducible components,
\begin{equation}
Q_\mnr \ = \ Q^{\rm (tr1)}{}_{\mnr} \ + \ Q^{\rm (tr2)}{}_\mnr \ + \ Q^{\rm (s)}{}_\mnr \ + \ Q^{\rm (tn)}{}_\mnr,
\end{equation}
where the first term is the trace of the non-metricity one-form $\dfQ_{\nu\rho} = Q_\mnr \dex x^\mu$,  the second one is the rest of the trace, $Q^{\rm (s)}{}_\mnr$ is the totally symmetric part without trace, and $Q^{\rm (tn)}{}_\mnr$ is the trace-free tensor component with no totally symmetric part ($Q^{\rm (tn)}{}_{(\mnr)} = 0$). Defining
\begin{equation}
Q^{(1)}{}_\mu \ = \ Q_{\mu\sigma}{}^\sigma, \hsp{2cm}
Q^{(2)}{}_\mu \ = \  Q^\sigma{}_{\sigma\mu},
\end{equation}
the different components can be expressed as
\bea
Q^{\rm (tr1)}{}_\mnr &=& \frac{1}{\dimM} \, Q^{(1)}{}_\mu \, g_{\nu\rho} \,,\\
Q^{\rm (tr2)}{}_\mnr &=& \frac{2}{(\dimM-1)(\dimM+2)}
\left[ \frac{1}{\dimM}g_{\nu\rho}Q^{(1)}{}_{\mu} \, - \, g_{\mu(\nu}Q^{(1)}{}_{\rho)} \, - \, g_{\nu\rho}Q^{(2)}{}_{\mu} \, + \, \dimM g_{\mu(\nu}Q^{(2)}{}_{\rho)}\right]\,,\\
Q^{\rm (s)}{}_{\mnr} &=& Q_{(\mu\nu\rho)} \ -\ \frac{1}{\dimM+2} \, g_{(\mu\nu}  \left(Q^{(1)}{}_{\rho)}\, +\, 2Q^{(2)}{}_{\rho)}\right)\,,\\
Q^{\rm (tn)}{}_{\mnr}&=& Q_{\mu\nu\rho}\ - \ Q^{\rm (tr1)}{}_\mnr\ - \ Q^{\rm (tr2)}{}_\mnr \ - \ Q^{\rm (s)}{}_{\mnr} \,.
\eea
The $ \half \dimM^2 (\dimM+1)$ independent components of the full non-metricity are distributed over
its irreducible parts as follows:  each of the traces, $Q^{\rm (tr1)}{}_\mnr$ and $Q^{\rm (tr2)}{}_\mnr$ has $\dimM$ independent components, $Q^{\rm (s)}{}_{\mnr}$ has $\tfrac{1}{6}\dimM (\dimM-1)(\dimM+4)$ and the remaining $\tfrac{1}{3}\dimM (\dimM^2-4)$ constitute the irreducible part $Q^{\rm (tn)}{}_{\mnr}$.

%%%%%%%%%%%%%%%%%%%%%%%%%%%%%%%%%%%%%%%%%%%%%%%%%
\section{Brief review of differential forms}
\label{AppendixDiffForms}
The exterior notation is very natural when dealing with gauge theories. It is known that metric-affine gravity can be seen as a gauge theory of the $\dimM$-dimensional affine group \cite{HMMN}. In fact, the basic physical objects in this formalism are differential forms over the spacetime with values in certain representation of the gauge group (depending on how the latin indices transform),
\begin{equation}
\dfal^{a...b} \ =\ \frac{1}{k!}\, \alpha_{\mu_1 ... \mu_k}{}^{ a...b} \, \dex x^{\mu_1} \wedge ...\wedge \dex x^{\mu_k}\,.
\end{equation}
For example, the metric $g_{ab}$ and the coframe $\cofr^a$ transform homogeneously (in the tensor product representation of the fundamental one), however the connection $\dfom_a{}^b$ transforms under the adjoint representation.

\subsubsection*{Interior product}

The \emph{interior product} by a vector, $\dint{\boldsymbol{V}} = V^a\dint{\boldsymbol{e}_{a}} = V^\mu \dint{\boldsymbol{\partial}_{\mu}}$, is the linear operator that acts on $1$-forms as
\begin{equation}
  \dint{\boldsymbol{\partial}_{\mu}}\dfal = \alpha_\mu \qquad \Leftrightarrow \qquad \dint{\boldsymbol{e}_{a}}\dfal = e^\mu{}_a \alpha_\mu \,,
\end{equation}
which implies for example $\dint{\boldsymbol{e}_{a}}\cofr^{b}=\delta_a^b$ or $\dint{\boldsymbol{e}_{a}}\dex x^\mu= e^\mu{}_a$. This operation is extended to forms of arbitrary rank by imposing the graded Leibniz rule
\begin{equation}
  \dint{\boldsymbol{e}_{a}}(\dfal \wedge \dfbe) \ = \ (\dint{\boldsymbol{e}_{a}}\dfal)\wedge \dfbe \ + \ (-1)^p \dfal\wedge(\dint{\boldsymbol{e}_{a}}\dfbe) \,,
\end{equation}
where $p=\mathrm{rank}(\dfal)$. In particular, for a general $p$-form:
\begin{align}
  \dint{\boldsymbol{e}_{a}}\dfal 
  &= \ \frac{1}{(p-1)!} e^\nu{}_a \alpha_{\nu \mu_1 ... \mu_{p-1}} \dex x^{\mu_1}\wedge ...\wedge \dex x^{\mu_{p-1}} \nonumber \\
  &= \ \frac{1}{(p-1)!} \alpha_{a b_1 ... b_{p-1}} \cofr^{b_1} \wedge ...\wedge \cofr^{b_{p-1}} \,.
\end{align}

\subsubsection*{Hodge duality}

A metric structure in a manifold induces an isomorphism between the space of $p$-forms and the space of $(\dimM-p)$-forms (for each $p$). This isomorphism, called the \emph{Hodge duality}, can be explicitly given by the Hodge star operator:
  \begin{align}
    \star\,:\,\Omega^{p}(M) & \longrightarrow\Omega^{\dimM-p}(M) \nonumber \\
    \dfal \quad & \longmapsto \quad \star\dfal \ = \ \frac{1}{(\dimM-p)!p!}\, \alpha^{b_1...b_p} \, \LCtensor_{b_1...b_p c_1...c_{\dimM-p}}\, \cofr^{c_1}\wedge ...\wedge\cofr^{c_{\dimM-p}}\,.
  \end{align}
We have omitted the possible external indices of $\dfal$, since this isomorphism does not affect them. In the definition of the Hodge star we have introduced the Levi-Civita tensor\footnote{Note that when we omit the indices of the determinant, we always refer to the determinant in the coordinate basis, $g\ = \det(g_{\mu \nu})$. It should not be confused with $\det(g_{ab})$. For that reason we write them explicitly in this expression.}
\begin{equation}
\LCtensor_{a_1 \dots a_\dimM} = \sqrt{|\det(g_{ab})|} \ \LCsymbol_{a_1 \dots a_\dimM}\label{DefLCTensor} \,, 
\end{equation}
where $\LCsymbol_{a_1 \dots a_\dimM} = \dimM! \delta^1_{[a_1} ... \delta^\dimM_{a_\dimM]}$ is the $\dimM$-dimensional alternating symbol. Two important properties of the Levi-Civita tensor, which we will use often in our calculation are the following,
\begin{align}
  \Dex \LCtensor_{a_1...a_\dimM} \ &= \ -\frac{1}{2}\, \LCtensor_{a_1...a_\dimM}\, \dfQ_c{}^c \,,  \label{DexLCtensor} \\
  \LCtensor_{a_1...a_k c_1...c_{\dimM-k}}\LCtensor^{b_1...b_k c_1...c_{\dimM-k}} \ &= \ \mathrm{sgn}(g) k! (\dimM-k)! \delta^{b_1...b_k}_{a_1...a_k}\,, \label{contractionLCtensor}
\end{align}
where $\Dex$ is the exterior covariant derivative defined in (\ref{extcovder}) for an arbitrary connection
 $\dfom_a{}^b$.

%%%%%%%%%%%%%%%%%%%%%%%%%%%%%%
\section{Metric-affine equations of motion for general curvature dependent Lagrangians}
\label{AppendixEoM}

In metric-affine gravity, the Noether identities of $\mathrm{Diff}(M)$ and $\mathrm{GL}(\dimM,\mathbb{R})$ imply that only the equations of motion of $\cofr^a$ and $\dfom_a{}^b$ are necessary, since the equation of the metric is identically satisfied if the other two are \cite{HMMN}. 

Consider a Lagrangian that depends on the connection only through the curvature, i.e.
\begin{equation} 
  S[g,\cofr,\dfom] = \int \dfL(g_{ab},\cofr^a,\dfR_{a}{}^{b}(\dfom)) = \int \mathcal{L}(g_{ab},e^a{}_{\mu},R_{\mu \nu a}{}^{b}(\dfom)) \sqrt{|g|} \dex^\dimM x \,,
\end{equation}
The equations of the coframe and the connection can be expressed in the language of differential forms as
\bea
  0=\frac{\delta S}{\delta\cofr^{a}}  & =& \frac{\partial\dfL}{\partial\cofr^{a}} 
   \ = \ \dint{\boldsymbol{e}_{a}}\dfL \ - \ (\dint{\boldsymbol{e}_{a}}\dfR_{c}{}^{b})\wedge\left(\frac{\partial\dfL}{\partial\dfR_{c}{}^{b}}\right)\,, \label{GeneralEoMCofr} \\ [.2cm]
  0=\frac{\delta S}{\delta\dfom_{a}{}^{b}}  & =& \Dex\left(\frac{\partial\dfL}{\partial\dfR_{a}{}^{b}}\right) \label{GeneralEoMConn} \,,
\eea
where the variation and the partial derivative with respect to differential forms have been defined extracting the variations from the left:
\begin{equation} 
  \delta S[\dfal] = \int \delta \dfal \wedge\frac{\delta S}{\delta \dfal} \,, \qquad{}
  \delta \dfL (\dfal, \dex \dfal,...) = \delta \dfal \wedge\frac{\partial \dfL}{\partial \dfal} + \delta\dex\dfal \wedge\frac{\partial \dfL}{\partial \dex\dfal}+...    
\end{equation}
Equivalently, in tensor notation we would obtain
\begin{align}
  0\ =\ \frac{1}{\sqrt{|g|}}\frac{\delta S}{\delta e^{a}{}_{\mu}} 
    & \ = \ e^{\mu}{}_{a} \mathcal{L} + \frac{\partial\mathcal{L}}{\partial e^{a}{}_{\mu}}\,,\\
  0\ = \ \frac{-1}{2\sqrt{|g|}}\frac{\delta S}{\delta\omega_{\mu a}{}^{b}} 
    & \ = \ \left(\nabla_{\lambda}- \frac{1}{2}Q_{\lambda \sigma}{}^{\sigma}+T_{\lambda\sigma}{}^{\sigma}\right) \left(\frac{\partial\mathcal{L}}{\partial R_{\lambda\mu a}{}^{b}}\right) \ - \ \frac{1}{2}T_{\lambda\sigma}{}^{\mu} \frac{\partial\mathcal{L}}{\partial R_{\lambda\sigma a}{}^{b}}\,. 
\end{align}

%%%%%%%%%%%%%
\subsubsection*{Useful particular case}

Suppose we apply the  following  splitting of the connection
\begin{equation} 
  \dfom_{ab} \ = \ \bar \dfom_{ab} \ + \ \half \, \dfQTrLess  _{ab} \ + \ \tfrac{1}{2\dimM}\,\dfQ_c{}^c g_{ab} \,,
\end{equation}
and that the remaining theory is both independent of $\dfQ_c{}^c$ and $\bar\Dex \dfQTrLess_{ab}$, such that we have an action of the type
\begin{equation} 
  \hat{S}[g,\cofr,\bar\dfom,\dfQTrLess] = \int \hat\dfL(g_{ab},\cofr^a,\bar\dfR_{a}{}^{b}(\bar\dfom),\dfQTrLess_{ab}) \,.
\end{equation}
In that case, one can prove that the equation of motion of the new variables $\bar\dfom_a{}^b$ and $\dfQTrLess_{ab}$ are:
\bea
  0=\frac{\delta \hat{S}}{\delta\dfQTrLess_{ab}}        & =& \frac{\partial\hat\dfL}{\partial\dfQTrLess_{ab}} \,,\label{GeneralEoMTrLessQ}  \\ [.2cm]
  0=\frac{\delta \hat{S}}{\delta \bar\dfom_{a}{}^{b}}  & = & \bar\Dex\left(\frac{\partial\hat\dfL}{\partial\bar\dfR_{a}{}^{b}}\right) \,. \label{GeneralEoMbarConn}
\eea

%%%%%%%%%%%%%%%%%%%%%%%%%%%%%%%%%%%%%%%%%%%%%%%%%%%%%%%%%%
\end{document}